\documentclass[%
 reprint,
superscriptaddress,
 amsmath,amssymb,
 aps,
prl
]{revtex4-2}

\usepackage{changes}
\usepackage{graphicx}
\usepackage{dcolumn}
\usepackage{bm}
\usepackage[colorlinks=true,linkcolor=blue,citecolor=blue,urlcolor=blue]{hyperref}

\begin{document}

\preprint{APS/123-QED}

\title{The fate of quasiparticles at high-temperature}

\author{A. Hunter}
\affiliation{Department of Quantum Matter Physics, University of Geneva, 24 Quai Ernest-Ansermet, 1211 Geneva 4, Switzerland}
 
\author{S. Beck}
\affiliation{Center for Computational Quantum Physics, Flatiron Institute, 162 Fifth Avenue, New York, NY 10010, USA}

\author{E. Cappelli}
\affiliation{Department of Quantum Matter Physics, University of Geneva, 24 Quai Ernest-Ansermet, 1211 Geneva 4, Switzerland}

\author{F. Margot}
\affiliation{Department of Quantum Matter Physics, University of Geneva, 24 Quai Ernest-Ansermet, 1211 Geneva 4, Switzerland}

\author{M. Straub}
\affiliation{Department of Quantum Matter Physics, University of Geneva, 24 Quai Ernest-Ansermet, 1211 Geneva 4, Switzerland}

\author{Y. Alexanian}
\affiliation{Department of Quantum Matter Physics, University of Geneva, 24 Quai Ernest-Ansermet, 1211 Geneva 4, Switzerland}

\author{G. Gatti}
\affiliation{Department of Quantum Matter Physics, University of Geneva, 24 Quai Ernest-Ansermet, 1211 Geneva 4, Switzerland}

\author{M. D. Watson} 
\affiliation{Diamond Light Source, Harwell Campus, Didcot, OX11 0DE, United Kingdom} 

\author{T. K. Kim} 
\affiliation{Diamond Light Source, Harwell Campus, Didcot, OX11 0DE, United Kingdom} 

\author{C. Cacho} 
\affiliation{Diamond Light Source, Harwell Campus, Didcot, OX11 0DE, United Kingdom} 

\author{N.C. Plumb}
\affiliation{Swiss Light Source, Paul Scherrer Institut, CH-5232 Villigen PSI, Switzerland}

\author{M. Shi}
\affiliation{Swiss Light Source, Paul Scherrer Institut, CH-5232 Villigen PSI, Switzerland}

\author{M. Radovi\'{c}}
\affiliation{Swiss Light Source, Paul Scherrer Institut, CH-5232 Villigen PSI, Switzerland}

\author{D. A. Sokolov}
\affiliation{Max Planck Institute for Chemical Physics of Solids, Dresden, Germany}
\author{A. P. Mackenzie}

\affiliation{Max Planck Institute for Chemical Physics of Solids, Dresden, Germany}
\affiliation{Scottish Universities Physics Alliance, School of Physics and Astronomy, University of St. Andrews, St. Andrews KY16 9SS, United Kingdom}
\author{M. Zingl}

\affiliation{Center for Computational Quantum Physics, Flatiron Institute, 162 Fifth Avenue, New York, NY 10010, USA}

\author{J. Mravlje}
\affiliation{Department of Theoretical Physics, Institute Jozef Stefan, Jamova 39, SI-1001 Ljubljana, Slovenia}

\author{A. Georges}
\affiliation{Center for Computational Quantum Physics, Flatiron Institute, 162 Fifth Avenue, New York, NY 10010, USA}

\affiliation{Coll\`{e}ge de France, 11 Place Marcelin Berthelot, 75005 Paris, France}
\affiliation{PHT, CNRS, \'{E}cole Polytechnique, IP Paris, F-91128 Palaiseau, France}

\author{F. Baumberger}
\affiliation{Department of Quantum Matter Physics, University of Geneva, 24 Quai Ernest-Ansermet, 1211 Geneva 4, Switzerland}
\affiliation{Swiss Light Source, Paul Scherrer Institut, CH-5232 Villigen PSI, Switzerland}

\author{A. Tamai}
\affiliation{Department of Quantum Matter Physics, University of Geneva, 24 Quai Ernest-Ansermet, 1211 Geneva 4, Switzerland}

\newcommand{\SRO}{Sr$_2$RuO$_4$}
\newcommand{\ReS}{$\Sigma^{\prime}$}
\newcommand{\ImS}{$\Sigma^{\prime\prime}$}

\date{\today}

\begin{abstract}

We study the temperature evolution of quasiparticles in the correlated metal \SRO. Our angle resolved photoemission data show that quasiparticles persist up to temperatures above 200~K, far beyond the Fermi liquid regime.
Extracting the quasiparticle self-energy 
we demonstrate that the quasiparticle residue $Z$ increases with increasing temperature.
Quasiparticles eventually disappear on approaching the bad metal state of \SRO{} not by losing weight but via excessive broadening from super-Planckian scattering.
We further show that the Fermi surface of \SRO{} -- defined as the loci where the spectral function peaks -- deflates with increasing temperature. These findings
are in semi-quantitative agreement with dynamical mean field theory calculations.
\end{abstract}

\maketitle

Many correlated electron systems with diverse magnetic and electronic ground states turn into bad metals at high temperature -- that is their resistivity increases with temperature and shows no sign of saturation well beyond the Mott-Ioffe-Regel (MIR) limit where the mean free path defined in semiclassical transport models drops below interatomic distances~\cite{Emery1995,Gunnarsson2003,Hussey2004}.
Examples include cuprates~\cite{Takagi1992}, ruthenates~\cite{Tyler1998}, iron-pnictides~\cite{Kamihara2008,Si2008}, manganites~\cite{Takenaka2002}, alkali doped $C_{60}$~\cite{Hebard1993}, and organic salts~\cite{Limelette2003}.

A pragmatic definition of a quasiparticle, adopted throughout this article, is the existence of a clearly discernible peak in the spectral function.
The bad metal state at or beyond the MIR limit does not host quasiparticle-like excitations because with a coherence length below the Fermi wave length, a particle like description is no longer appropriate.
This is qualitatively consistent with ARPES studies of cobaltates~\cite{Valla2002}, ruthenates~\cite{Damascelli2001,Wang2004,Kondo2016} or iron-chalcogenides~\cite{Yi2015,Rhodes2017}, which all found that QPs disappear at temperatures well below the MIR limit. This behavior was thus far interpreted as a gradually decreasing QP residue $Z$ with increasing temperature~\cite{Valla2002,Wang2004}. 
A crossover where $Z\to 0$ with increasing temperature is also found in a slave-boson~\cite{Mezio2017} and in dynamical mean field theory (DMFT)~\cite{Limelette2003,Merino2000} studies of the single band Hubbard model for undoped systems close to the metal-insulator transition (MIT). Work on organic salts found that the transition $Z\to 0$ manifests itself as a crossover from a bad metal regime in which $Z$ is finite to a semiconductor-like resistivity at high temperature where $Z\approx 0$~\cite{Limelette2003}. Such a resistive transition is reminiscent of the c-axis resistivity in \SRO~\cite{Hussey1998} but is not observed for in-plane transport in \SRO{} and most other bad metals.

An alternative picture that is intuitively appealing for metallic systems is that at elevated temperatures excitations become more bare-electron like and that QPs become short-lived but simultaneously lose renormalization such that $Z\rightarrow1$.
This behavior is indeed found in DMFT studies of the doped Hubbard model~\cite{Deng2013,Xu2013} and of Hund metals~\cite{Mravlje2011,Deng2016,Kugler2020}.

The QP residue $Z$ is defined as 
$Z_k = \left(1\left.-\partial\Sigma^\prime(\omega,k)/\partial\omega\right|_{\omega = 0}\right)^{-1}$ 
where $\Sigma^\prime$ is the real part of the self-energy. $Z$ corresponds to the integral of the coherent QP peak in the spectral function and is thus also called QP weight. However in real systems there is no established way to distinguish the ’coherent’ from the ’incoherent’ part of the spectrum and therefore extracting $Z$ from an analysis of spectral weight may be difficult. Moreover in ARPES data it is often not obvious how to separate increasingly broad peaks from the background which is often poorly understood.
\begin{figure*}[t]
\centering
\includegraphics[width=1\textwidth]{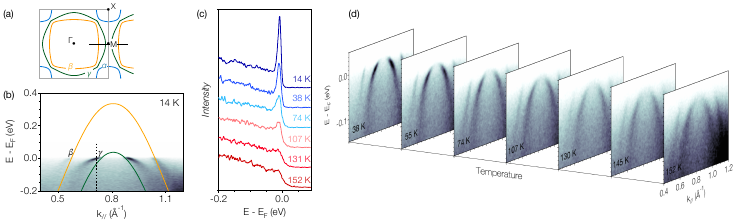}
\caption{(a) Schematic of the Fermi surface of \SRO. (b) Low-temperature ARPES data showing the $\beta$ and $\gamma$ band dispersing symmetrically around the Brillouin zone boundary. Yellow and green lines are the bare dispersions of the $\beta$ and $\gamma$ bands obtained from the Hamiltonian introduced in Ref.~\cite{Tamai2019}. (c) EDCs extracted for the $\gamma$ band at the momentum indicated by a dashed line in (b), corresponding to a QP energy of $-10$~meV.(d) Temperature dependence of the ARPES spectral functions for the same cut shown in (b).}
\label{fig:img_plots} 
\end{figure*}

Here, we address these issues by studying the fate of QPs in \SRO{} with increasing temperature. Analyzing the dispersion measured by ARPES rather than the spectral weight, we show that the QP residue $Z$ increases in \SRO{} with increasing temperature, contrary to earlier reports. QPs eventually disappear at elevated temperature by 'dissolving' in the incoherent part of the spectrum rather than by losing spectral weight. 
We further show that the evolution of $Z$ with temperature derived from our data is in agreement with DMFT calculations and fully consistent with the ARPES spectral weight.

\SRO{} is an ideal material to address the evolution of the QP residue $Z$ with temperature. 
In its normal state, \SRO{} is a prototypical correlated metal with a well-understood electronic structure, a large mass enhancement and sharp QP peaks at low-temperature~\cite{Mackenzie1996,Damascelli2000,Tamai2019}. Resistivity data show a well defined Fermi liquid regime below $T_{FL}\approx 25$~K crossing over to an extended regime with T-linear resistivity~\cite{Maeno1997,Tyler1998}. Depending on the precise criterion, the MIR limit is surpassed for $\rho=0.1 - 0.7$~m$\Omega$cm reached at $T= 300 - 800$~K~\cite{Tyler1998}. Crucially for our approach here, the self-energy of \SRO{} is dominated by local electron-electron correlations and thus does not have significant momentum dependence~\cite{Shen2007,Tamai2019}. In this limit the QP residue is related to the velocity renormalization \mbox{$\frac{v_F}{v_{\textrm{bare}}}=Z$} where $v_F$ and $v_{\textrm{bare}}$ are the QP and bare Fermi velocity, respectively. 
We will base our quantitative analysis on this relation 
(and refer to this quantity as 'QP residue')
before demonstrating that $Z$ obtained in this way is fully consistent with 
the integral of coherent spectra (referred to as 'QP weight').

Fig.~1 shows ARPES data of \SRO{} acquired at a photon energy of 40~eV. Details of the measurement conditions are given in Supplemental Material~\footnote{See Supplemental Material, which includes Refs.~\cite{Blaha2001,Marzari/Vanderbilt:1997,Souza/Marzari/Vanderbilt:2001,PARCOLLET2015398,aichhorn_dfttools_2016,Kim2018,Seth2016274,Yates_et_al:2007,Chmaissem:1998,bulla08,zitko09,NRGLjubljana,Zingl2019}, for experimental and theoretical methods, data analysis and a discussion of the temperature evolution of Z in the half-filled single orbital Hubbard model.}.
Throughout this paper we focus on the momentum space cut marked by a thick black line in Fig.~1(a). Along this high-symmetry line, orbital hybridization is minimal and
the $\beta$ and $\gamma$ sheets retain $\approx 80\%$ $xz/yz$ and $xy$ orbital character, respectively~\cite{Tamai2019}. 

The comparison in Fig.~1(b) with the bare band dispersion introduced in Ref.~\cite{Tamai2019}  illustrates the strong and orbital dependent mass enhancement of the QP excitations in \SRO{} documented in the literature~\cite{Mackenzie1996, Bergemann2003,Tamai2019}.
Intriguingly, monitoring the spectral function with increasing temperature reveals an apparent dichotomy between energy distribution curves (EDCs) and the two-dimensional energy -- momentum images. The EDCs of the $\gamma$ band (Fig.~1(c)) show a well defined sharp peak at low-temperature that decays with increasing temperature and appears to be swallowed by the background around 130~K. Similar observations in earlier work were interpreted as a transition $Z\to 0$ and have been related to the crossover to semiconducting c-axis transport in \SRO~\cite{Hussey1998,Wang2004}. Interestingly though, the ARPES image plots in Fig.~1(d) show QP like bands up to higher temperature with no sign of a transition at 130~K. Additional data in Supplemental Material Fig.~9 shows that band like states persist up to $\approx 250$~K. This illustrates that analyzing spectral weights in EDCs is delicate and may not be robust.
We note that the apparent suppression of peaks in EDCs is not an artifact of the Fermi cutoff. In Fig.~1(c) we deliberately extracted EDCs at an initial state energy of $-10$~meV to minimize effects of the Fermi-Dirac distribution.

\begin{figure*}[htp]
\centering
\includegraphics[width=0.98\textwidth]{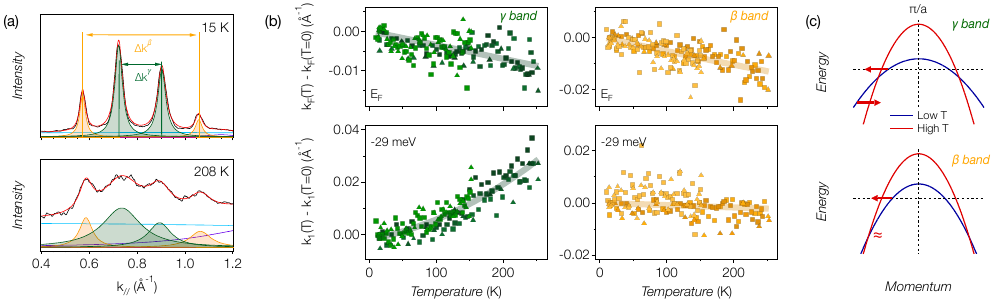}
\caption{(a) MDCs at $E_F$ extracted from ARPES data with Lorentzian fits to quantify peak positions of the $\beta$ and $\gamma$ band. Additional examples of fits are shown in Supplemental Material. (b) Temperature dependence of $k(T) - k(T=0)$ at $E = E_F$ and $E = -29$~meV. Different colors and symbols are used for different experimental runs and temperature up- and down-sweeps.
(c) Schematic illustrating the temperature evolution of the QP dispersion of the $\beta$ and $\gamma$ band.}
\label{fig:dks} 
\end{figure*}

We start our quantitative analysis by fitting the peak positions of the $\beta$ and $\gamma$ bands in momentum distribution curves (MDCs). Assuming that both bands disperse symmetrically around the Brillouin zone boundary, we obtain their Fermi wave vectors from $k_F^{\beta,\gamma} = \pi/a - \Delta k^{\beta,\gamma}/2$, where $\Delta k^{\beta,\gamma}$ is the separation of the Fermi level crossings in the first and second Brillouin zone, as indicated in Fig.~2(a). We note that finite energy resolution leads to an 'up turn' in the apparent dispersion measured by ARPES with peak positions in MDCs that deviate from the quasiparticle dispersion~\cite{Levy2014}. We have quantified this artifact (see Supplemental Material) using DMFT spectral functions -- which are known to provide a good description of high-resolution laser-ARPES data~\cite{Tamai2019} -- and show in Fig.~2(b) directly the corrected wave vectors.
As a further cross-check we performed 2D fits of the ARPES data that include resolution effects directly as a convolution with a 2D response function. The results from these 2D fits are fully consistent with the MDC analysis shown here (see Supplemental Material). 

Intriguingly, our analysis shows that $k_F$ shrinks with increasing temperature for both bands. This signifies a deflation of both the $\beta$ and $\gamma$ sheets. Such a change in Fermi surface volume cannot arise from inter-orbital charge transfer. Neither does it imply a reduction of the carrier density. Our DMFT calculations -- in which the integral over the spectral function precisely reproduces the number of electrons --
show a comparable deflation of the Fermi surfaces, as shown in Supplemental Material, Fig.~6. This suggests that the temperature evolution of the Fermi surfaces is an intrinsic property of \SRO.
Changes of the FS size upon heating can occur when particle-hole symmetry is broken and have been studied theoretically in the single band doped Hubbard model~\cite{Deng2013,Xu2013,Osborne2021}. 
In this case a transition from hole-like to electron-like at high temperature can even occur~\cite{Deng2013,Xu2013}.

Extending the analysis of MDCs to energies below the chemical potential shows a remarkable orbital differentiation. The QP band position $k_1$ at $E - E_F = -29$~meV increases markedly for the $\gamma$ sheet, while it is nearly temperature independent for the $\beta$ sheet. 
We have verified that this behavior cannot be attributed to the effects of thermal expansion on the bare band structure.
Hence, we conclude that both QP bands 'un-renormalize' with increasing temperature but the $\gamma$ sheet does so much more rapidly and in a slightly different way, as indicated pictorially in Fig.~2(c). 

\begin{figure*}[htp]
\centering
\includegraphics[width=\textwidth]{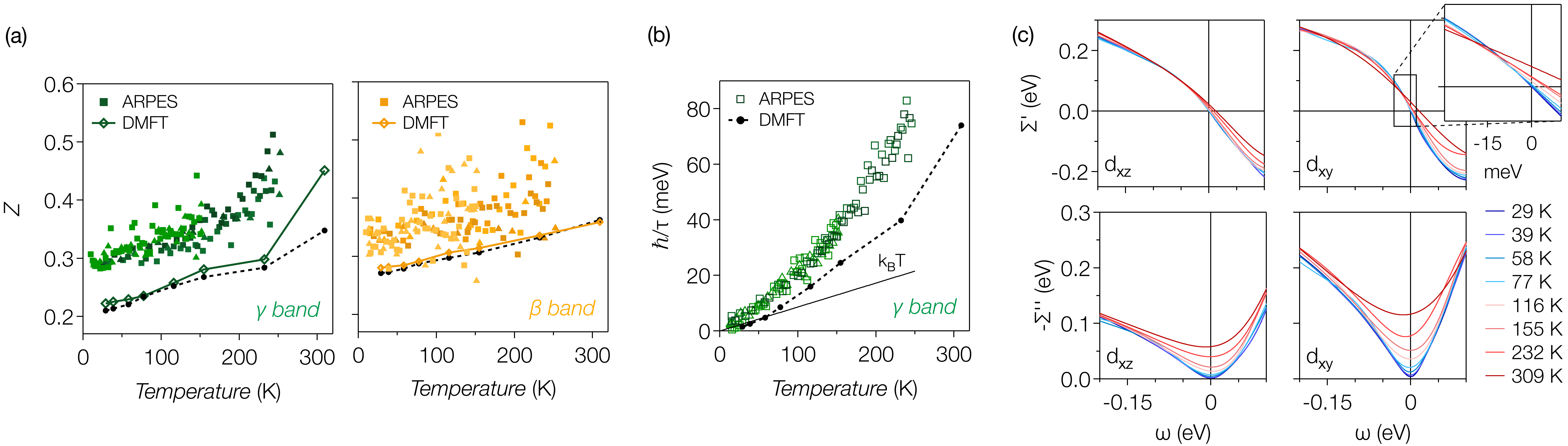}
\caption{(a) Temperature dependence of $Z=v_F/v_{\textrm{bare}}$ for the $\gamma$ and $\beta$ bands (filled symbols) compared to DMFT calculations. The dashed black lines 
give $Z_{\nu}=\left(\sum Z_m^{-1}|U_{m\nu}(k)|^2\right)^{-1}$.
Diamond symbols connected by lines are obtained by repeating the ARPES analysis on DMFT spectral functions.
(b) QP line width $\Gamma_{\textrm{QP}}=\hbar/\tau$ (full width half maximum) obtained from 2D fits compared with DMFT results and the Planckian limit $k_B T$.
In (a) and (b) different symbols and colors are used for different experimental runs and temperature up- and down-sweeps. (c) DMFT self-energies.}
\label{fig:Z_dmft} 
\end{figure*}

We estimate the QP residue $Z$ from our analysis by assuming a parabolic dispersion of the $\beta$ and $\gamma$ bands symmetrically around the Brillouin zone boundary at $\pi/a$. In this case, the QP Fermi velocity $v_F (T)$ is uniquely determined by $k_F (T)$ and $k_1 (T)$. Fig.~3(a) shows $Z(T)=v_F(T)/v_{\textrm{bare}}$ obtained in this approximation for $v_{\textrm{bare}}^{\beta}=2.57$~eV\AA{} and $v_{\textrm{bare}}^{\gamma}=1.06$~eV\AA{} obtained from the bare Hamiltonian in Ref.~\cite{Tamai2019}. 
We find that $Z$ in \SRO{} increases with temperature.
For a direct comparison of our experimental results with DMFT calculations we transform $Z^{\textrm{DMFT}}$ from an orbital to a band basis, as described in Ref.~\cite{Tamai2019} (dashed black lines in Fig.~3(a)). This reproduces the slope of $Z(T)$ found in experiment. We note that the QP residues obtained from our MDC analysis are slightly higher than found in DMFT and in a previous ARPES study restricted to low temperatures~\cite{Tamai2019}. 
This difference is reduced in a 2D analysis of the experimental data (see Supplemental Material).

The temperature dependence of $Z$ affects the QP scattering rate $\Gamma_{QP}=\hbar/\tau$. In Fig.~3(b) we compare $\Gamma_{\textrm{QP}}$ obtained from 2D fits of the experimental data with DMFT and the Planckian value $k_{B}T$~\cite{Bruin2013, Hartnoll2022}.
We restrict this comparison to the $\gamma$ band, for which the 2D fits are more reliable.
The DMFT QP line width is calculated from $\Gamma_{\nu}=\sum \Gamma_m|U_{m\nu}(k)|^2$ where $m,\nu$ are the orbital and band indices, respectively, $\Gamma_m = -\Sigma^{\prime\prime}(\omega,T)\left(1-\frac{\partial \Sigma^{\prime}(\omega,T)}{\partial\omega}|_{\omega} \right)^{-1}$ and $|U_{m\nu}|^2$ is the orbital content obtained from the bare Hamiltonian.
One sees that the scattering rate exceeds the Planckian bound already for $T\gtrsim 50$~K and continues to grow with a positive curvature. 
The scattering rates obtained from the ARPES data are in quantitative agreement with resistivity measurements (see Supplemental Material).
We further note that the slight underestimation of $\hbar/\tau$ in DMFT is fully consistent with the small contribution of electron-phonon scattering identified in recent theoretical work~\cite{Abramovitch2023}.

Fig.~3(c) shows that the deflation of the Fermi surfaces of \SRO{} and the 'un-renormalization' of its QP dispersion reported here are readily discernible in the DMFT self-energies. The slope of \ReS{} reduces for both orbitals with increasing temperature, implying that the Fermi fluid in \SRO{} is less renormalized at higher temperature. At the same time \ReS$(T,\omega=0)$ increases with temperature causing a positive QP energy at the bare Fermi wave vector and thus a deflation of the Fermi surface.

\begin{figure}[b]
\centering
\includegraphics[width=0.95\columnwidth]{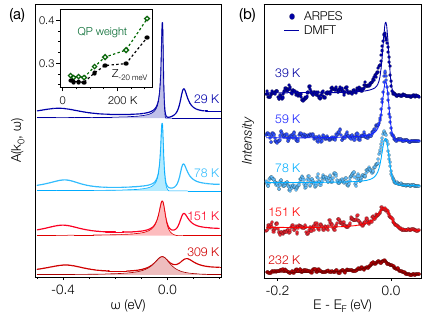}
\caption{(a) DMFT spectral functions calculated for a momentum where the QP energy of the $\gamma$ band is $\approx -20$~meV (near the dashed black line in Fig.~1(b)). Shaded areas indicate the coherent QP spectral weight as identified by fits (see main text). The inset shows the temperature dependence of the QP weights (green) compared to $Z(\omega=-20$~meV) extracted from $\Sigma^\prime$ (black). 
(b) Background subtracted ARPES EDCs of the $\gamma$ band QP at -10~meV
compared to corresponding DMFT spectral functions multiplied by a Fermi function and convolved by a Gaussian to account for the experimental resolution.}
\label{fig:QP_weight} 
\end{figure}

We now return to the evolution of spectral weights with temperature. 
Our findings in Fig.~3 of $Z$ which increases with temperature indicates that the QP spectral weight should also increase with temperature.  However, this appears to be in contradiction with the temperature dependent EDCs presented in Fig.~1~(c) which show QP peaks gradually becoming swallowed by the background.  Indeed, it is this picture of temperature dependent EDCs which has led to previous conclusions of $Z$ decreasing as temperature is increased~\cite{Wang2004}.  This apparent discrepancy poses an important conceptual question: are QP weights and velocity normalizations responding differently to a change in temperature?  To answer this question, we first analyze QP weights in DMFT spectral functions.

Fig.~4(a) shows DMFT spectra $A(k_0,\omega)$ calculated for a momentum $k_0$ where the QP energy of the $\gamma$ band is $\approx-20$~meV. The main peak just below the chemical potential can thus be attributed to the $\gamma$ band. Additional peaks in the unoccupied states and around $-0.4$~eV originate from the $\beta$ and $\alpha$ bands, respectively. As expected, the $\gamma$ band QP peak broadens progressively with increasing temperature and its height diminishes. At the same time, it evolves from the typical asymmetric Fermi liquid line shape at low-temperatures towards a Lorentzian 
at elevated temperature.

To estimate the QP spectral weight we expand the self-energy to first order around the QP peak position $\omega_0$:
$\Sigma^{\prime}=\Sigma^{\prime}_0+(1-1/Z)\delta\omega$
and $\Sigma^{\prime\prime}=\Sigma^{\prime\prime}_0+\alpha\:\delta\omega$, where $\delta\omega = \omega-\omega_0$. The spectral function is then given by
    $A(k,\delta\omega) \approx \frac{Z}{\pi}\frac{Z (\Sigma^{\prime\prime}_0+\alpha\:\delta\omega)}{\delta\omega^2 + Z(\Sigma^{\prime\prime}_0+\alpha\:\delta\omega)^2}$.
Fits of the DMFT spectra with this expression show that the first order approximation is sufficiently flexible to capture the evolution of the line shape with temperature and provides a physically meaningful estimate of the QP weight.
Significantly, we find that the QP weight does increase with temperature despite the peaks becoming less intense.
The direct comparison of the QP weight from these fits with $Z(\omega_0)=(1-\frac{\partial\Sigma^{\prime}}{\partial w}|_{\omega_0})^{-1}$ obtained from the DMFT self-energy (inset of Fig.~4(a)) shows minor quantitative differences but illustrates that the concept of QP weight remains meaningful at elevated temperature in DMFT spectra of multiband systems.

We now address the consistency of ARPES spectral weights with the increase in $Z$ found from the ARPES dispersions. To this end, we first isolate the intrinsic spectral function from the ARPES spectra by subtracting a "background". The latter arises from inelastically scattered electrons but may also contain contributions from the interference of multiple photoemission channels and is notoriously hard to model~\cite{Kevan1992,Miller1996}. 
We therefore resort to a pragmatic approach and approximate the background with the spectrum at $\pi/a$ where no direct transitions are observed in the relevant energy range (see Supplemental Material, section VIII for more details).
In contrast to the approach of Ref.~\cite{Wang2004}, this results in spectra with QP like excitations persisting up to the highest temperatures of $\approx 250$~K studied in our experiment (Fig.~4(b)).

Direct fits of these background subtracted spectra with Eq.~1 proved to be unstable. 
However, the evolution of the line shape and peak intensity in our experimental data is in excellent agreement with DMFT spectral functions as shown in Fig.~4(b). Note that in this comparison we only apply a global, temperature-independent scaling factor between the DMFT and the experimental spectra. This implies that the experimental spectral weights are fully consistent with an increase in $Z$ with increasing temperature.

In summary, our work shows that QPs in \SRO{} are resilient up to temperatures approaching the MIR limit.
Notably, we find no abrupt changes in behaviour as the temperature crosses $T_{FL} \approx 25$~K or through the metal-insulator crossover in $\rho_c$ at $\approx 130$~K.  
Quantitative analysis shows that the QP residue $Z$ increases with increasing temperature. We expect this behavior to be common in bad metals. However, it is not necessarily universal. Hund metals with an orbital selective Mott phase, as proposed by Yi~\textit{et al.} for certain iron-chalcogenides, may show QP weights $\to 0$ for orbitals that localize while the weights of other orbitals remain finite or even increase with temperature~\cite{Yi2015}.
Our findings are likely to have relevance to a broad range of correlated materials.
Intriguingly, a recent ARPES study of the Mott insulator V$_2$O$_3$ reported an increasing spectral weight on approaching the metal-insulator transition by cooling from the metallic high-temperature phase~\cite{Thees2021}, opposite to the behavior found here for the Hund metal \SRO. Yet, it is presently not known whether this observation is consistent with the QP dispersion in V$_2$O$_3$ and little is known about the temperature dependence of the QP residue in other Mott systems including cuprates and iridates.

\begin{acknowledgements}
We thank J. Schmalian and P.D.C. King for discussions.
The experimental work was supported by the Swiss National Science Foundation (SNSF). We acknowledge Diamond Light Source for time on Beamline I05 under Proposal SI25083 and the Paul Scherrer Institut, Villigen, Switzerland for  provision of synchrotron radiation beamtime at the SIS beamline of the SLS.
The Flatiron Institute is a division of the Simons Foundation. 
\end{acknowledgements}

\providecommand{\noopsort}[1]{}\providecommand{\singleletter}[1]{#1}%

\end{document}